\documentclass[conference]{IEEEtran}
\IEEEoverridecommandlockouts

\usepackage[utf8]{inputenc}
\usepackage[T1]{fontenc}
\usepackage{cite}
\usepackage{amsmath,amssymb,amsfonts}
\usepackage{algorithm,algorithmic}
\usepackage{graphicx}
\usepackage{hhline}
\usepackage{textcomp}
\usepackage{xcolor}
\def\BibTeX{{\rm B\kern-.05em{\sc i\kern-.025em b}\kern-.08em
    T\kern-.1667em\lower.7ex\hbox{E}\kern-.125emX}}

\usepackage[hyphens]{url}

\usepackage{times}

\usepackage{optidef}
\usepackage{caption}
\usepackage{nohyperref}

\begin{document}
\title {\textbf {\Large A User-experience Driven SSIM-Aware Adaptation Approach for DASH Video Streaming\\}}

\author{\IEEEauthorblockN{Mustafa Othman}
\IEEEauthorblockA{\textit{L2TI, Galilee Institute}\\
		University Paris 13 \\
Villetaneuse, France \\
Email: mustafa.othman@univ-paris13.fr}
\and
\IEEEauthorblockN{ Ken Chen}
\IEEEauthorblockA{\textit{L2TI, Galilee Institute}\\
		University Paris 13 \\
Villetaneuse, France \\
Email: ken.chen@univ-paris13.fr}
\and
\IEEEauthorblockN{Anissa Mokraoui}
\IEEEauthorblockA{\textit{L2TI, Galilee Institute}\\
		University Paris 13 \\
Villetaneuse, France \\
Email: anissa.mokraoui@univ-paris13.fr}
}

\def\kctmp#1{ {\bf KC {\sl #1}} }

\def\algosigle {SBA}
\def\algoname {SSIM Based Adaptation}


\maketitle
\begin{abstract}
Dynamic Adaptive Streaming over HTTP (DASH) is a video streaming technique largely used. One key point is the adaptation mechanism which resides at the client's side. This mechanism impacts greatly on the overall Quality of Experience (QoE) of the video streaming. In this paper, we propose a new adaptation algorithm for DASH, namely \algoname~(\algosigle). This mechanism is user-experience driven: it uses the Structural Similarity Index Measurement (SSIM) as main video perceptual quality indicator; moreover, the adaptation is based on a joint consideration of SSIM indicator and the physical resources (buffer occupancy, bandwidth) in order to minimize the buffer starvation ({\em rebuffering}) and video quality instability, as well as to maximize the overall video quality (through SSIM). To evaluate the performance of our proposal, we carried out trace-driven emulation with real traffic traces (captured in real mobile network). Comparisons with some representative algorithms (BBA, FESTIVE, OSMF) through major QoE metrics show that our adaptation algorithm \algosigle~achieves an efficient adaptation minimizing both the rebuffering and instability, whereas the displayed video is maintained at a high level of bitrate.
\end{abstract}

\section{Introduction}
\label{sec: inter in}

\begin{figure}[htbp]
	\centering%
	\fbox{\includegraphics[width=0.8\columnwidth]{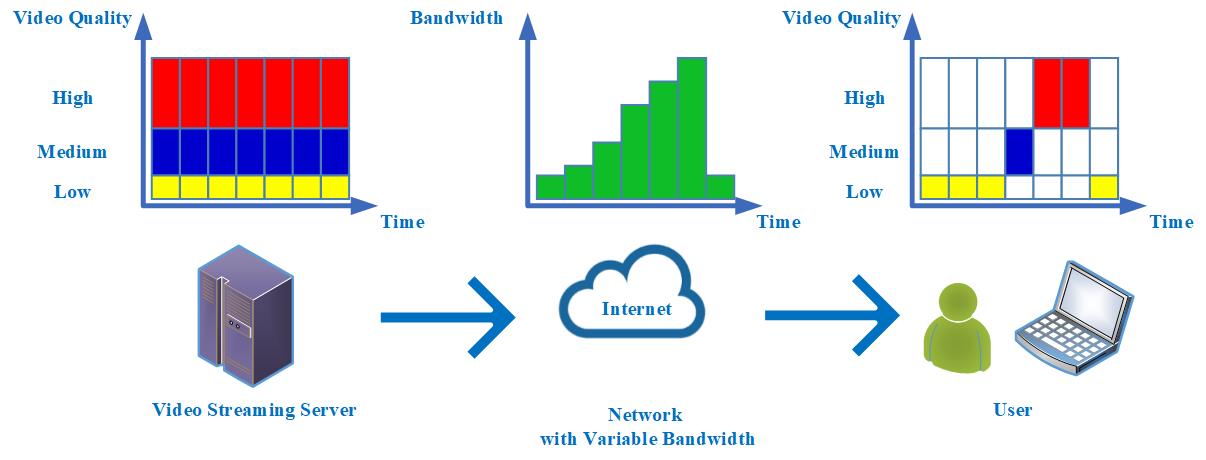}}
	\caption{DASH Streaming Flow Process.}
	\label{fig:DASH}
\end{figure}

Recent studies predicted that the growth of mobile traffic would take up to 20\% of total Internet traffic by 2021 \cite{networking2016forecast}. Moreover, it is expected that video traffic will reach 80\% of all internet traffic by 2021. The majority of video streaming on the internet today uses the MPEG’s Dynamic Adaptive Streaming over HTTP (DASH) standard which aims to deliver video with high Quality of Experience
(QoE) \cite{nam2016qoe, seufert2015survey, dobrian2011understanding}.
The principle of DASH consists in dividing the entire video into segments, called {\em chunk}, in order to send them separately by using HTTP. Each chunk has several versions, each one is encoded with a specific bitrate. Chunks are fetched by clients through HTTP/TCP. This approach makes DASH popular~\cite{wang2017has} \cite{hartung2011drm} since a) it can be built above the omni-present HTTP and b) client can easily choose, for each chunk, the bitrate which is most suitable to the current network conditions with some Adaptive Bitrate (ABR) algorithm \cite{huang2015buffer, zhou2013buffer, jiang2014improving, akhshabi2012happens, cicalo2014quality} in order to maximize the user’s QoE. Figure \ref{fig:DASH}  depicts the DASH streaming process. Today, major content providers (including NETFLIX and YouTube) use DASH.


%

Among the main challenges related to DASH scheme~\cite{dobrian2011understanding} \cite{krishnan2013video} \cite{mok2011inferring}, there are in particular:
\begin{enumerate}
	\item {\em The rebuffering}: this term refers to the freezing of video playback when the buffer is empty and waiting for the next video chunk. The duration and the frequency of rebuffering during a video streaming session is among the most important metrics that affect the user's QoE.
	\item {\em The instability}: it is considered as another important QoE metric. When the bitrate changes from one chunk to the next one, there is surely some variations on the perceived video quality. If the quality of the video among consecutive segments is too abrupt and/or too frequent, the user's QoE will be bad.
\end{enumerate}

It is a real challenge to  minimize the rebuffering and instability, and at the same time to maximize
the overall quality of the received video. Actually, the surest way to get best stability and avoid rebuffering
is to send video always at the lowest bitrate level. However, we would have the worst video quality. 
So, the selection of the level (in terms of bitrate) of the next chunk is a matter of balance
between QoE metrics such as the rebuffering, the instability and the video quality.

Video quality can be assessed through several objective metrics, among which Structural Similarity Index Measurement~(SSIM) and Peak Signal to Noise Ratio~(PSNR). The PSNR suffers from its inconsistency with the human eye perception video quality. The SSIM is considered to be able to better capture
the difference between the original and the encoded images and  provides a measurement which is  closer to what is visually noticeable as defaults by a human being \cite{cicalo2014quality} \cite{georgopoulos2013towards} \cite{pudlewski2012compressed}.

This paper proposes a new adaptation algorithm for DASH, namely \algoname~(\algosigle), by using the SSIM indicator to select the quality of the next video chunk with the objectives of optimizing the QoE by maximizing the video quality and minimizing the rebuffering and instability. To achieve such a balance, the main idea resides in the fact that we decide to a) {\em increase} the bitrate level only when the SSIM indicates a significant improvement in the video quality (thus getting more video content at almost the same {\em user perceived} video quality), and b) 
{\em decrease} the bitrate level only when there is a real risk of rebuffering (thus minimize the instability). 
 
Our proposal consists in adding the SSIM values for each level of video chunks in Media Presentation Description (MPD), which is the standard way in DASH to provide clients useful informations for video adaptation. Our main contribution is to explore this additional information (SSIM value), in combination with the classical ones, in order to achieve a better adaptation.  
 
\begin{figure}[htbp]
	\centering%
	\fbox{\includegraphics[width=0.8\columnwidth]{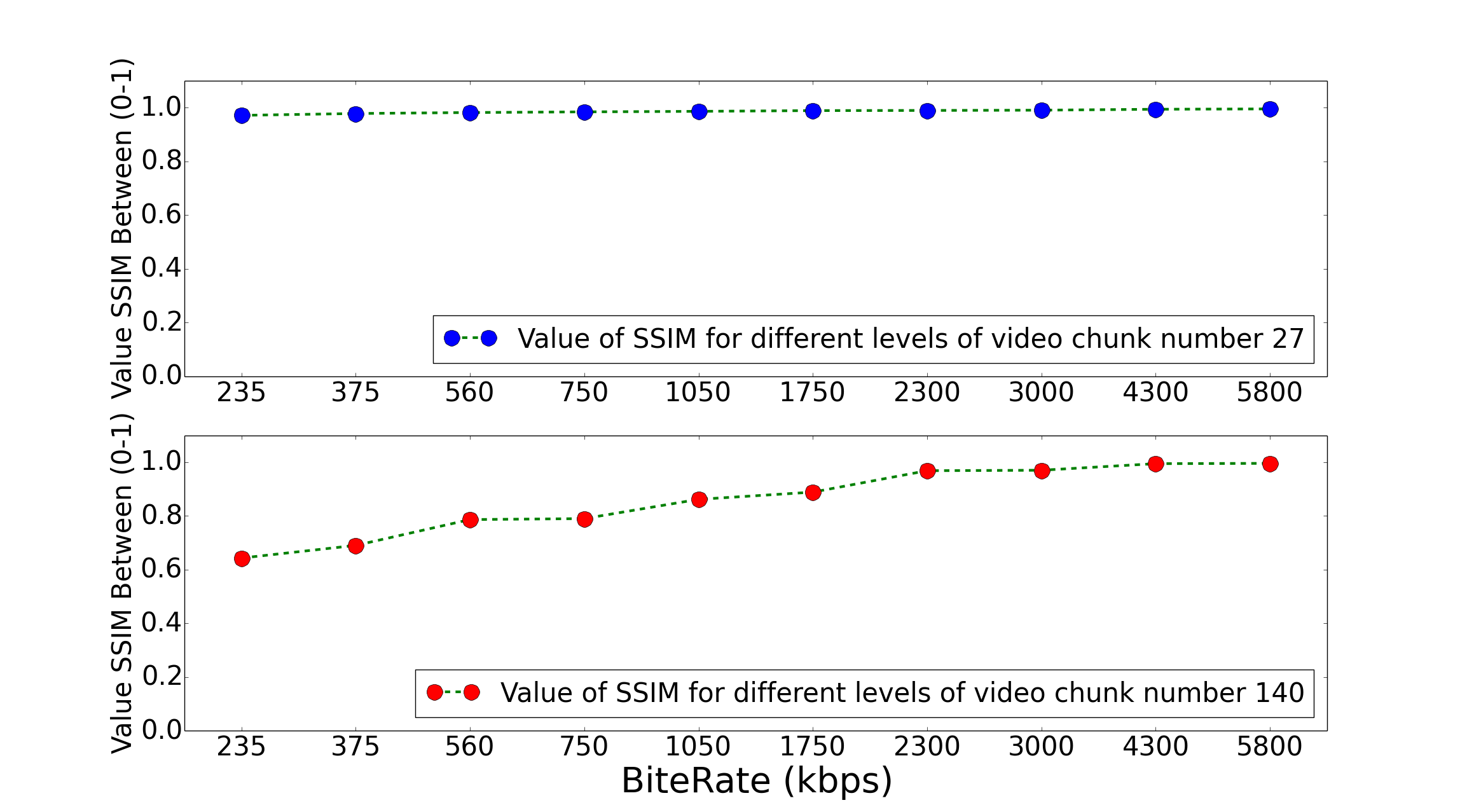}}
	\caption{Values of SSIM for different resolution for the chunck number 27 and 140.}
	\label{fig:SSIM for Chunk nomber 27 and 140}
\end{figure}

Figure \ref{fig:SSIM for Chunk nomber 27 and 140} helps to better illustrate our basic idea. In this figure, each point gives the bitrate of one of the levels. It can be observed that for chunck number 27, all the levels offer nearly the same SSIM value; whereas  for chunck number 140, the lifting in SSIM value for higher levels (hier bitrates) are rather noticeable. This leads to the idea of including the SSIM value among criteria for level selection. Indeed, it is not efficient to select a higher bitrate level when the lower level offers a very comparable video quality.

The rest of this paper is organized as follows. Section \ref{sec:state-of-art} provides the state of the art on the bitrate adaptation algorithms problem and video quality metrics. Section \ref{sec:algo} presents our algorithm. Section \ref{sec:evaluation} compares the results of our algorithm with a selection of relevant algorithms in the literature. Section \ref{sec:conclusion} concludes the paper.

\section{Related Work}
\label{sec:state-of-art}
Bitrate adaptation algorithms are generally classified into three categories:
\begin{enumerate}
	\item {\em Buffer-based}: the buffer occupancy is used as the main indicator for the selection of the bitrate (level) of the next  video chunk to be downloaded
	\cite{huang2015buffer} \cite{zhou2013buffer}.
	\item {\em Rate-based}:  the bitrate (level) of the next video chunk to be downloaded is chosen to maximize the use of the (estimated) future bandwidth \cite{jiang2014improving} \cite{akhshabi2012happens}.
	\item Mixture:  a mix of the two previous categories \cite{cicalo2014quality}.
\end{enumerate}

Huang et al. \cite{huang2015buffer} proposed the Buffer Based Adaptation (BBA) method, which aims to:
a) Avoid unnecessary rebuffering and b) Maximize the overall video quality. They used the {\em buffer occupancy} as a control signal to select the level (bitrate) of the next video chunk instead of the estimated bandwidth. 
They calculate dynamically a couple of <maximum, minimum> bitrate levels. When possible, the actual video rate keeps growing to the maximum, until there is not enough (with respect to a threshold value) room in buffer, then the video bitrate drops to the minimum level. 

Jiang et al. \cite{jiang2014improving} proposed a general bitrate adaptation algorithm framework called Fair, Efficient, and Stable adapTIVE algorithm~(FESTIVE). This adaptation algorithm  aims to improve fairness, stability and efficiency of the DASH player by predicting throughput to be the harmonic mean of the experienced throughput for the past 5 downloaded chunks, as well as a stability score as a function of the bitrate switches in the past 5 chunks. As the  prediction does not take into account the buffer occupancy, FESTIVE may have rebuffering which affects the QoE performance.

The Open Source Media Framework (OSMF) \cite{riad2015channel} is an HTTP video streaming platform developed by Adobe Systems. OSMF adaptation algorithm selects the next level of video chunk based on the chunk download duration. However, it could lead to instability because of the variability and unpredictability of the bandwidth especially in mobile networks.

Much work has been developed to evaluate the quality of the perceived visual information.  Several objective and subjective metrics were proposed with or without reference.  
Most of them aim to get as close as possible to the human visual system in order to apprehend the different visible distortions (induced by compression, contrast enhancement or transmission...) affecting  the original visual information. Among the different metrics, we retain the Structural Similarity Index Measurement (SSIM), considered as an objective metric \cite{wang2004video}, due to its proven performance. The state of the art shows that this metric has also been selected as a metric to assess the QoE (see e.g., ~\cite{cicalo2014quality} \cite{georgopoulos2013towards} \cite{pudlewski2012compressed} \cite{zinner2010impact}). 

In \cite{zinner2010impact}, the authors  rely on objective QoE metrics as SSIM which is considered as a full reference model. This allows conducting extensive measurement studies and deriving simple relationships applicable to QoE control.
They took into account the kind of the videos (e.g., interview, soccer match, movie).
Their results showed that they can use SSIM to determine the QoE behaviour of different content, they also shown that video sequences with lower resolution perform better than video sequences with a lower frame-rate.


In \cite{georgopoulos2013towards}, the authors  incorporated the SSIM into their {\em Utility Function} as the quality metric. The rational can be summarized as follows. First, the relationship between bitrate and perceptual quality is not linear; as the bitrate increases, the gain in video quality is gradually saturated. Second, the equal division of network bandwidth for video streams of different resolutions (i.e., a vertical line representing a certain bitrate) results in unfair video quality levels as perceived by end-users. In \cite{pudlewski2012compressed}, the authors used the SSIM to measure the quality of video transmission through their  system (Compressive Distortion Minimizing Rate Control, C-DMRC). The latter uses a  distributed cross-layer control algorithm that aims to maximize the received video quality over a multi-hop wireless network with lossy links.

\section{Proposed Adaptation Algorithm}
\label{sec:algo}
This section presents our new adaptation algorithm,  referred subsequently as \algoname~(\algosigle).

\subsection{Rationale}
Our algorithm combines the networking level control (buffer-based and rate-based) and the SSIM video quality metric. The key point consists in using the SSIM indicator to determine the level (bitrate and so video quality) of the next chunk to be fetched. A bitrate upgrade takes place not only because it is allowed by the network, but also because it would provide a real gain of {\em user perceived} quality. The algorithm tries to achieve the following goals~:
\begin{itemize}
	\item By estimating the available bandwidth, we always try to get the best achievable quality.
	\item We upgrade to a higher bitrate level only when there is a real gain in quality. Thus, 
	we try to maximize the video content downloading (and so minimize rebuffering) at (almost) the same video quality.
	\item We use buffer-occupancy as rebuffering alert signal and decrease 	the bitrate, 
	similar to the TCP's behaviour, only when there is a real risk. In this way, we minimize the instability.
\end{itemize}

The outline of the proposed algorithm is summarized as follows.
\begin{itemize}
	\item At the beginning of the streaming, as the networking situation is not known yet,  the algorithm starts at the lowest bitrate level.
	\item As the streaming goes on, the algorithm  gets a better estimation of the available bandwidth (noted subsequently as $EBW$).
	\item The algorithm upgrades to a higher bitrate level not only  when it is allowed by $EBW$, but also because this level provides a real gain on the video quality according to the SSIM indicator.
	\item In the case where the current $EBW$ is lower than the currently chosen video bitrate level, the algorithm chooses to not decrease the video to a lower bitrate.  In this way, the algorithm aims to maximize video quality and to minimize instability due to the bitrate level change. Of course, if the situation persists, there is a risk of buffer starvation.
	\item The algorithm gives priority to avoiding rebuffering:  a critical zone is defined in the buffer. Each time the buffer occupancy falls into this zone,  the algorithm decreases the bitrate level to the lowest one.
\end{itemize}

\subsection{Notation and Conditions}
The video is divided into $K$ chunks (video segments) where each chunk has an equal duration of $T$ seconds.
The video is encoded at $R$ bitrate levels, denoted as ${\cal R} =\{r_j\}_{j=1\dots R}$
i.e., each chunk has $R$ encoded versions respectively at $r_1$ to $r_R$ bitrate. By convention, the bitrate levels are ordered as follows  $r_1<r_2<\dots <r_R$.

For each chunk and at each bitrate level, the corresponding SSIM metric is computed as follows. The SSIM of each image of the chunk is first computed, then the SSIM of a chunk is deduced as the mean value of the SSIM of the different images of this chunk. The SSIM matrix of the video stream is then computed. It is composed of $Q(i,j)$ element corresponding to the SSIM of the $i$-th chunk at the $r_j$ bitrate level. From the implementation's viewpoint, this is compatible with the generic DASH framework. Indeed, this SSIM matrix can be pre-computed and stored in the server. It is then sufficient to incorporate it into the MPD (Media Presentation Description) so that the client can get it.

At the client-side, the algorithm estimates the available bandwidth with the following process:
\begin{itemize}
	\item Each time the client sends the fetch order of a chunk, say chunk $i$, this time is memorized as $s_i$. Upon the complete download of the current chunk at $t_i$ with an actual volume of $Vb_i$ bits, the bandwidth actually consumed by the download of chunk $i$ is computed as:
\begin{equation}\label{eqn:CBW}
CBW(i)=\frac{Vb_i}{t_i-s_i}.
\end{equation}
	\item The estimated bandwidth for the fetch of the $l$-th chunk ($l>1$), denoted by $EBW(l)$, is then computed as the mean average of the actually consumed bandwidth over the downloaded chunks:
\begin{equation}\label{eqn:EBW}
EBW(l)=\frac{\sum_{i=1}^{l-1} CBW(i)}{l-1}.
\end{equation}
\end{itemize}
For the first chunk, according to the proposed algorithm (as it will be explained later), the lowest bitrate version ($r_1$) will be used, i.e. $EBW_1=r_1$.

The algorithm keeps also the trace of the difference in SSIM between the adjacent chunks actually displayed.
Let $d_i$ be the level at which the chunk $i$ is displayed. For chunk $l$ ($l>1$), $\Delta(l)=Q(l, d_l)-Q(l-1, d_{l-1})$, is defined to measure the variation in terms of SSIM related to the previous chunk ($l-1)$ when the $l$-th chunk is displayed. The mean SSIM variation till chunk $l$ (for $l>1$), denoted by $\alpha(l)$, is then computed as below:
\begin{equation}\label{eqn:DefAlpha}
\alpha(l)= \frac{\sum_{k=2}^l \Delta(k)}{l-1}.
\end{equation}

Following the convention adopted by the scientific community, the buffer's capacity is given in seconds. The buffer has a capacity of BS (in seconds). It is divided into two regions, region C (for Critical) and region N (for Normal). When the buffer occupancy is below a threshold value (noted by $L_c$), we are in region C. The current buffer occupancy is denoted by $b$.

\subsection{Our Algorithm}

Hereafter, we describe the  \algosigle\ algorithm (cf. pseudo-code in Figure \ref{fig:algo1})
which aims to determine the bitrate level (denoted by $f$) of the next chunk  to be fetched.


\begin{figure}[htbp]
	\centering%
	\begin{algorithmic}[1]
		\renewcommand{\algorithmicrequire}{\textbf{Input:}}
		\renewcommand{\algorithmicensure}{\textbf{Output:}}
		\REQUIRE
		\STATE   $\mathcal{R}$, $Q$, $b$, $L_{c}$, $l$, $\alpha(l)$, $EBW(l)$, $d_{l-1}$, 
		\ENSURE  $f$: Bitrate level at which the next video chunk will be fetched
		\IF{$b$ <= $L_{c}$}
		\STATE {$f=r_1$}
		\ELSE
		\STATE {$p_l=\max \{r_j \in \mathcal{R}, r_j <EBW(l) \}$.}
		\IF{$\delta(l)=Q(l, p_l)-Q(l-1, d_{l-1})>\alpha(l)$}
		\STATE {$f=p_l$}
		\ELSE
		\STATE {$f=d_{l-1}$}
		\ENDIF
		\ENDIF
		\RETURN $f$ ;
	\end{algorithmic}		
	\caption{Adaptation algorithm.}
	\label{fig:algo1}
\end{figure}
This algorithm is run each time a fetch order can be issued, i.e. either when a chunk is totally displayed or a chunk is totally downloaded, and of course, when there is room in the buffer (i.e., $BS-b>T$). The parameters $\alpha(i)$ and $EBW(i)$ are assumed to be estimated through parallel processes.

Consider the fetch of the $l$-th chunk and denote the bitrate level of the previous chunk as $d_{l-1}$ and its SSIM $Q(l-1, d_{l-1}$). This algorithm has two regimes depending on the buffer occupancy:
\begin{enumerate} 
	\item When  the current buffer occupancy is  in the region C,  the next chunk will always be fetched at the lowest level, i.e. $f=r_1$.
	\item When the buffer occupancy is in the region N:
	\begin{enumerate} 
		\item With $EBW(l)$,  the {\em potential} bitrate $p_l$ which is the highest bitrate under $EBW(l)$ is determined.
		\item The difference in SSIM if $p_l$ {\em should be} used, 
		$\delta(l)=Q(l, p_l)-Q(l-1, d_{l-1})$, is then computed.
		\begin{enumerate} 
		\item If $\delta(l)>\alpha(l)$, the algorithm considers that there is sufficient gain in video quality and choose $p_l$ as the bitrate for chunk $l$, i.e. $f=p_l$.
		\item Otherwise, the next video chunk at the {\em current} level (i.e,. $f=d_{l-1}$) is fetched.
	\end{enumerate} 
	\end{enumerate} 
\end{enumerate} 

\section{Performance Evaluation}
\label{sec:evaluation}
This section compares and discusses the performance of the proposed SBA algorithm to a selection of competitive algorithms.

\subsection{Evaluation Framework}
To get a realistic networking context, a set of real traffic situation over the 4G mobile network of a major network provider has been collected. The traces were collected from different areas and periods in Paris to insure a large coverage of the traffic patterns. Three traditional test videos have been used:
\begin{enumerate}
	\item Animation (Big Buck Bunny) \cite{site01},
	\item Documentary film (Of Forests and Men) \cite{site05},
	\item Sport (The World’s Best Bouldering in Rocklands, South Africa) \cite{site06}.
\end{enumerate}
These videos are encoded with FFMPEG codec at the following levels (the ones used by Netflix)\cite{site02} \cite{site03}: ${\cal R}$ = $\{$235kbps, 375kbps, 560kbps, 750kbps, 1050kbps, 1750kbps, 2350kbps, 3000kbps, 4300kbps, 5800kbps$\}$.
Videos are then divided into chunks ($T=4$ seconds, \cite{riad2015channel}) by using MP4Box-GPAC framework \cite{site04}.

We developed (in Python) a simulator in order to evaluate the performance of DASH-based adaptation algorithms.  
This simulator can work in trace-driven mode, i.e., the networking context is reconstituted with real networking traces. 
The simulator reproduces timely the instants of chunk download completion (which depends on network condition) as well as the 
chunk playback (which can be blocked by rebuffering). At  each instant where the next chunk is to be downloaded, our algorithm enters in action by
computing the level of the next chunk. 

By using this simulator and the real-traffic trace previously mentioned, 
we compared SBA algorithm with the following three ones: BBA\cite{huang2015buffer}, FESTIVE\cite{jiang2014improving} and OSMF\cite{riad2015channel}. 
We have tested two scenarios with two different buffer sizes: a) $BS=120$ seconds, b) $BS=240$ seconds. 
Each scenario is tested with $24$ different traces. The threshold value ($Lc$) is 
set to $12$ seconds ($3$ chunks) in both scenarios.

The performance of the algorithms is assessed through 4 metrics (i.e., Rebuffering, Instability, SSIM, birate). For each metric, the average value is computed on $24$ tests:
\begin{enumerate}
	\item \textbf{Average Rebuffering}: is the average of  rebuffering (freezing) duration.
	\item \textbf{Average Instability}: is the average of bitrate changes.
	\item \textbf{Average of SSIM}: is the average of the SSIM of the video being displayed.
	\item \textbf{Average of bitrate}: is the average bitrate of the video being displayed.
\end{enumerate}

\subsection{Performance Analysis}
This section provides discussions on the achieved performance using the {\bf Animation} video stream. Table
\ref{tab:res-anim} summarizes the results for the two scenarios. For the first 4 lines in the table $BS=120$ s, whereas for the last 4 lines $BS=240$ s. One can observe that the SBA algorithm achieves the desired objective with shorter rebuffering, less instability at a good bitrate level.

\begin{figure}[htbp]
	\centering%
	\fbox{\includegraphics[width=0.8\columnwidth]{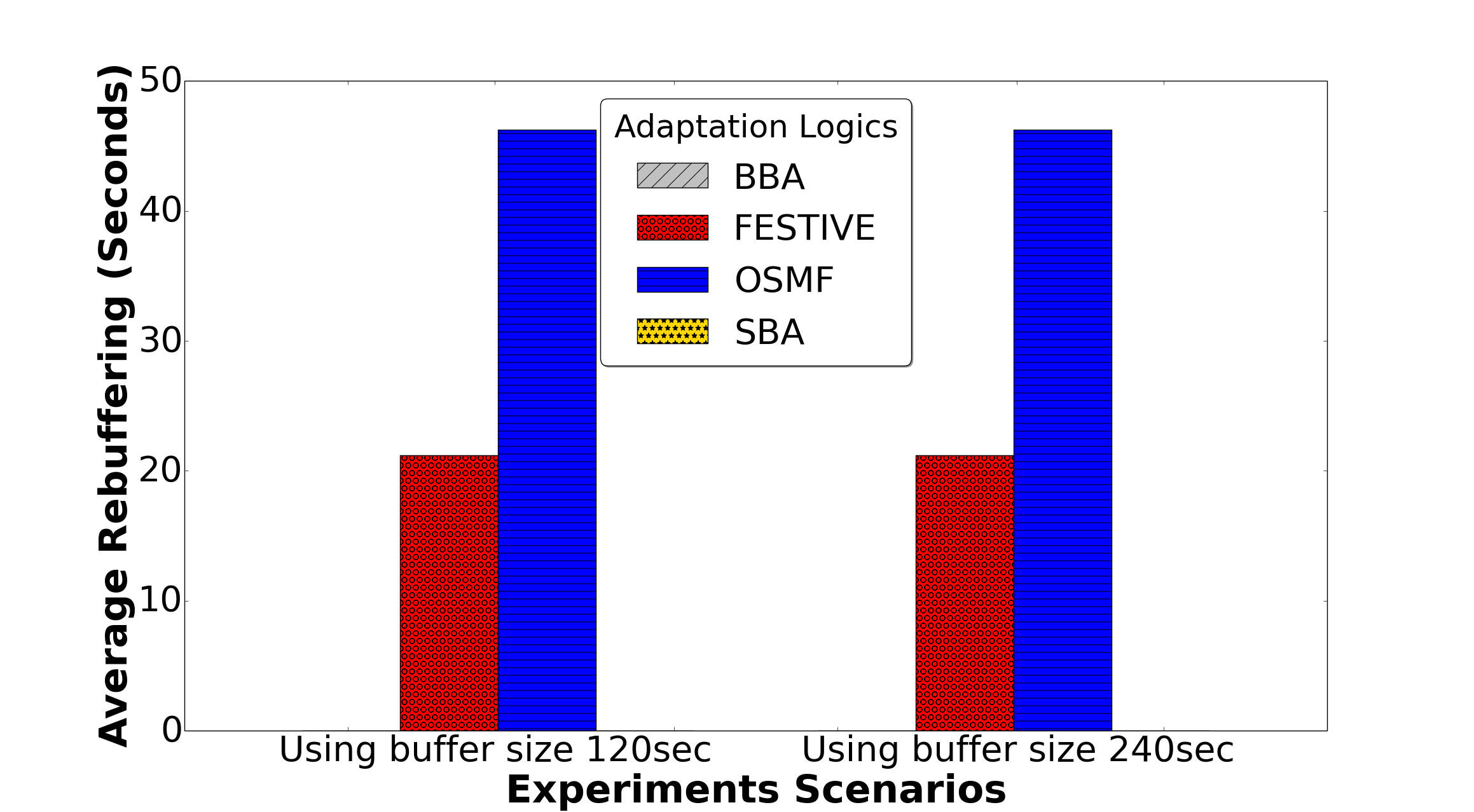}}
	\caption{Average Rebuffering duration for different algorithms with buffer sizes of $120$ and $240$ seconds and with animation (big buck bunny).}
	\label{fig:RebufDuration}
\end{figure}
Moreover, Figure \ref{fig:RebufDuration} shows that our proposal \algosigle~introduces zero rebuffering for both the scenarios. Actually, we give priority to rebuffering avoidance by setting a critical zone with drastic bitrate drop-off. Being a buffer-based algorithm, BBA works in a similar way and so  shows also the same zero rebuffering. On the contrary, FESTIVE and OSMF undergo rebuffering during video chunks playback for 21.208 and 46.25 seconds respectively. So, our algorithm performs better than FESTIVE and OSMF for the given scenarios.

\begin{figure}[htbp]
	\centering%
	\fbox{\includegraphics[width=0.8\columnwidth]{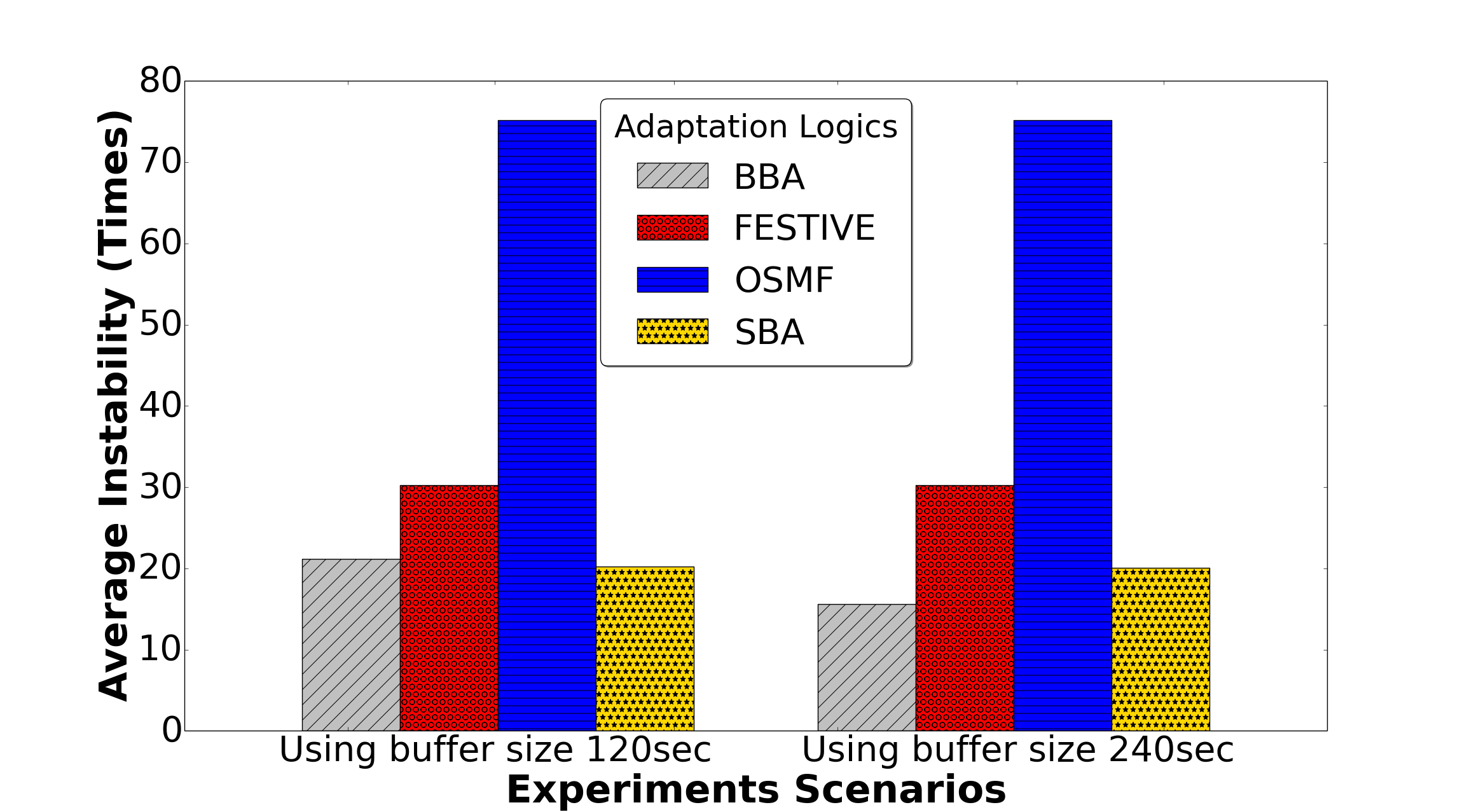}}
	\caption{Average Instability for different algorithms with buffer sizes of $120$ and $240$ seconds and with animation (big buck bunny).}
	\label{fig:Instability}
\end{figure}
\begin{figure}[htbp]
	\centering%
	\fbox{\includegraphics[width=0.8\columnwidth]{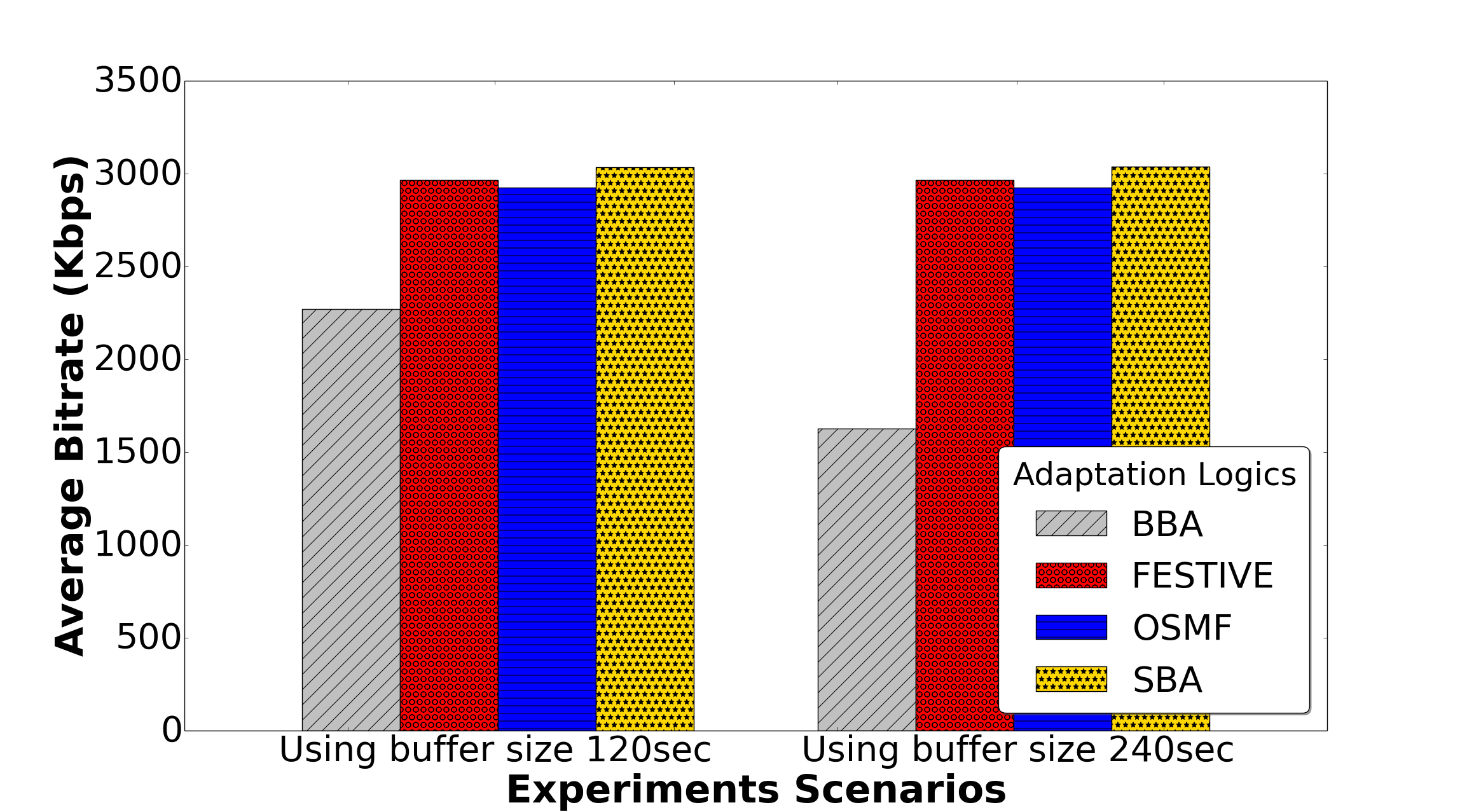}}
	\caption{Average Bitrate for different algorithms with buffer sizes of $120$ and $240$ seconds and with animation (big buck bunny).}
	\label{fig:bitrate}
\end{figure}

Figure \ref{fig:Instability} shows that our proposal \algosigle~achieves good performance, since it is respectively at the first (for $BS=120$ sec.) and second (for $BS=240$ sec.) places. For the scenario with $BS=240$ seconds, BBA algorithm is slightly better than our \algosigle: this is due to a more conservative bitrate 
increase approach of BBA. But the price to pay is a much lower average bitrate of BBA compared to the others, where as our algorithm keeps the highest average bitrate (cf. Figure \ref{fig:bitrate}).

\begin{figure}[htbp]
	\centering%
	\fbox{\includegraphics[width=0.8\columnwidth]{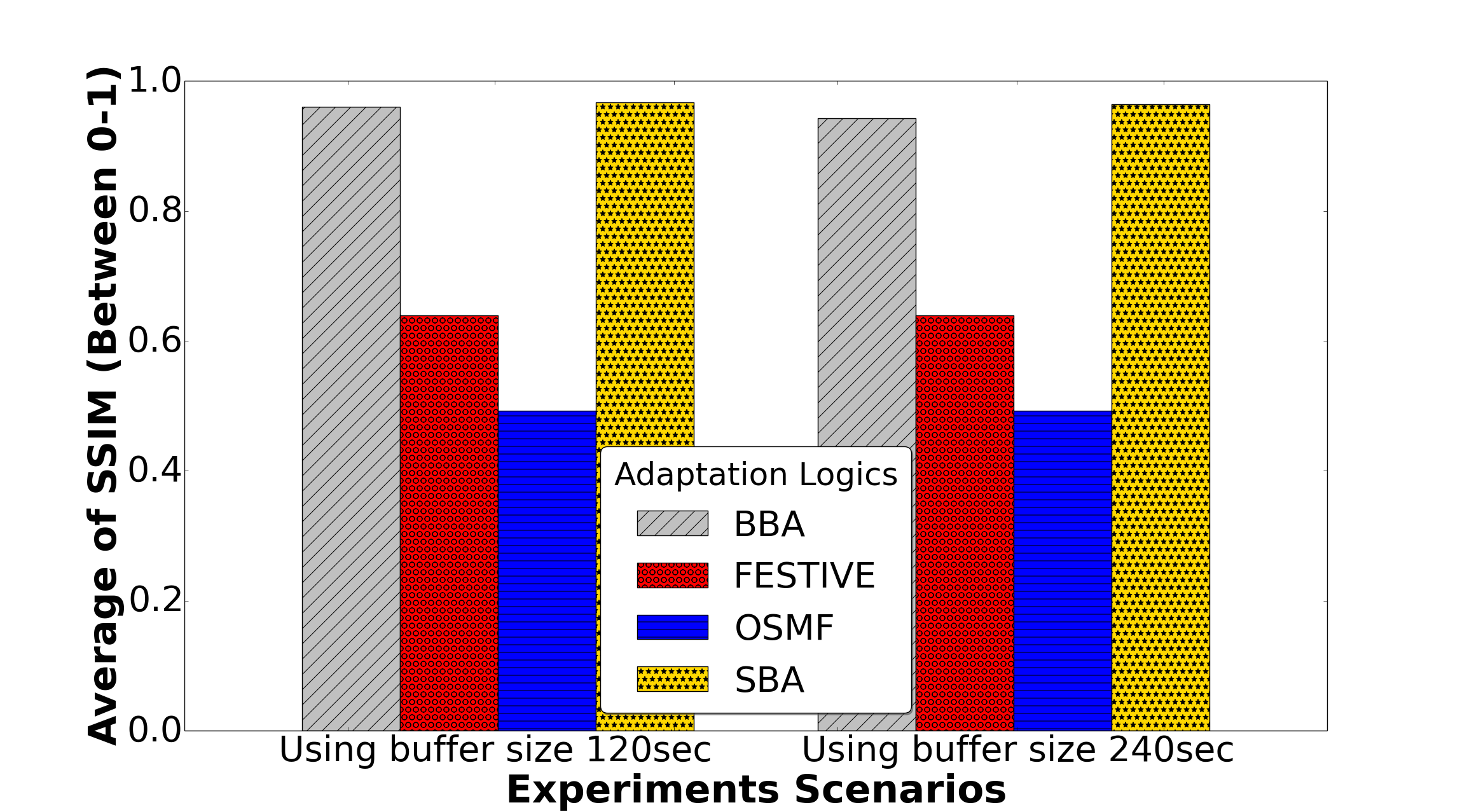}}
	\caption{Average SSIM for different algorithms with buffer sizes of $120$ and $240$ seconds and with animation (big buck bunny).}
	\label{fig:SSIM}
\end{figure}
As for the SSIM (see Figure \ref{fig:SSIM}),
our proposal \algosigle~and BBA have similar performance, which is much better than the two others.
This means in particular that our choice of upgrading only if there is a real gain in SSIM is justified.

Figure \ref{fig:bitrate} gives the \textbf{average bitrate} for different algorithms. As it is shown, our proposal \algosigle~achieves the highest average bitrate for both scenarios. We notices also that BBA, which have similar performance as our algorithm for the first 3 metrics, gets here the lowest bitrate,
probably because there is an excessive consideration for rebuffering avoidance.


\begin{table}[!t]
	\renewcommand{\arraystretch}{1.3}
	\caption{SUMMARIZED RESULTS OF THE TWO SCENARIOS WITH \textbf{ANIMATION}}
	\label{tab:res-anim}
	\centering
	\begin{tabular}{|c||c|c|c|c|}
		\hline
		\bfseries Adaptation Algo.  & \bfseries Rebuffering & \bfseries Instability & \bfseries SSIM & \bfseries BitRate\\ \hhline{|=#=|=|=|=|}
		SBA                         &\textbf{0.0}                &\textbf{20.25}	   &\textbf{0.967}	&\textbf{3035.165}\\
		BBA \cite{huang2015buffer}                         &\textbf{0.0}                & 21.166              &  0.960         &  2270.699\\
		FESTIVE \cite{jiang2014improving}                     &21.208                      &  30.208             &  0.638         &  2968.263 \\
		OSMF \cite{riad2015channel}                        &46.25                       &  75.166             &  0.492         & 2926.625 \\
		\hline
		\hline
		SBA                         &\textbf{0.0}	             &20.083               &\textbf{0.964}	&\textbf{3039.998}\\
		BBA \cite{huang2015buffer}                       &\textbf{0.0}                & \textbf{15.583}     &  0.942         &  1629.180\\
		FESTIVE \cite{jiang2014improving}                     &21.208                      & 30.208              &  0.638         &  2968.263 \\
		OSMF \cite{riad2015channel}                        &46.25                       &  75.166             &  0.492         &  2924.321\\
		\hline
	\end{tabular}
\end{table}

\subsection{Results Summary}
Additional results using the two test videos, namely {\bf Documentary} (see Table \ref{tab:res-doc}) and {\bf Sport} (see Table \ref{tab:res-sport}) are given in this section. Similar results to those obtained with {\bf Animation} stream are observed. One can notice that for both scenarios, our proposal \algosigle\ algorithm, achieves better ranking for most of the metrics.

\begin{table}[!t]
	\renewcommand{\arraystretch}{1.3}
	\caption{SUMMARIZED RESULTS OF THE TWO SCENARIOS WITH  \textbf{DOCUMENTARY}}
	\label{tab:res-doc}
	\centering
	\begin{tabular}{|c||c|c|c|c|}
		\hline
		\bfseries Adaptation Algo.  & \bfseries Rebuffering & \bfseries Instability & \bfseries SSIM & \bfseries BitRate\\   \hhline{|=#=|=|=|=|}
		SBA 					    &\textbf{0.0}		         &23.5				   &\textbf{0.941}	&2927.646\\
		BBA \cite{huang2015buffer}                    &\textbf{0.0} 		         &\textbf{19.916}	   &0.921		    &2326.034\\
		FESTIVE \cite{jiang2014improving}                     &22.625  			         &29.708			   &0.601			&\textbf{2963.134}\\
		OSMF \cite{riad2015channel} 	                    &44.625				         &76.958			   &0.459			&2848.442\\
		\hline
		\hline
		SBA 		  	            &\textbf{0.0}		         &23.416			   &\textbf{0.941}	&2929.382\\
		BBA \cite{huang2015buffer} 			            &\textbf{0.0} 		         &\textbf{14.458}	   &\textbf{0.941}	&1635.275\\
		FESTIVE \cite{jiang2014improving} 		            &22.625  			         &29.708			   &0.601			&\textbf{2963.134}\\
		OSMF \cite{riad2015channel}			            &44.625				         &76.958			   &0.459			&2841.568\\
		\hline
	\end{tabular}
\end{table}


\begin{table}[!t]
	\renewcommand{\arraystretch}{1.3}
	\caption{SUMMARIZED RESULTS OF THE TWO SCENARIOS WITH  \textbf{SPORT}}
	\label{tab:res-sport}
	\centering
	\begin{tabular}{|c||c|c|c|c|}
		\hline
		\bfseries Adaptation Algo.  & \bfseries Rebuffering & \bfseries Instability & \bfseries SSIM & \bfseries BitRate\\   \hhline{|=#=|=|=|=|}
		SBA 	                    &\textbf{0.0}  		         &\textbf{17.791} 	   &\textbf{0.957}	&\textbf{3086.453}\\
		BBA \cite{huang2015buffer} 	                    &\textbf{0.0}		         &17.875		       &0.952			&2470.888\\
		FESTIVE \cite{jiang2014improving}                     &18.541  	                 &28.0 		           &0.673			&2992.655\\
		OSMF \cite{riad2015channel} 	                    &46.625                      &77.083 		       &0.478			&2866.815\\
		\hline
		\hline
		SBA 		             	&\textbf{0.0}			     &17.75		           &\textbf{0.954}	&\textbf{3078.769}\\
		BBA \cite{huang2015buffer} 			            &\textbf{0.0}				 &\textbf{12.708}	   &0.929		    &1937.89\\
		FESTIVE \cite{jiang2014improving} 		            &18.541			             &28.0		           &0.673		    &2992.655\\
		OSMF \cite{riad2015channel}			            &46.625			             &77.083		       &0.478		    &2864.495\\
		\hline
	\end{tabular}
\end{table}

\section{Conclusion}
\label{sec:conclusion}
This paper proposed a new adaptation algorithm \algoname~(\algosigle) for DASH video streaming. 
This algorithm is user-experience driven since 
the main control factor is the Structural Similarity Index Measurement (SSIM) which 
is a good objective indicator for user perceived video quality.  
This algorithm takes jointly into consideration the networking level indicators 
(i.e., buffer occupancy, bandwidth) and
the SSIM to select the next level of video chunk. 
The performance analysis of the provided results carried out on trace-driven emulation with real traffic traces (captured in real mobile network) show that the proposed algorithm, compared to some representative algorithms (BBA, FESTIVE, OSMF) through major QoE metrics show that our algorithm (\algosigle) achieves a more efficient adaptation by minimizing both the rebuffering and instability, whereas the displayed video is maintained at a high level of bitrate.
Our main working direction being the joint consideration of networking mechanism and objective video quality metric, we continue in this direction for our future work. For SBA, we plan to explore the impact of the choice of the threshold value of SSIM. We also plan to extend this work to other relevant video quality metrics.


\section*{Acknowledgment}
{\footnotesize \textsuperscript{}}
\thanks{This work is supported in part by a scholarship of Campus France (878164H) as well as by a complementary support of the Galileo Graduate School-University Paris 13 (ED146).}

\bibliographystyle{IEEEtran}
\bibliography{source/citation}
\end{document}